# Ultrafast Dynamics of Defect-Assisted Auger process in $PdSe_2$ films: Synergistic Interaction Between Defect Trapping and Auger Effect

Di Li[1], Wenjie Zhang[1], Peng Suo[1], Jiaming Chen[1], Kaiwen Sun[1], Yuqin Zou[1], Hong Ma[2], Xian Lin[1], Xiaona Yan[1], Saifeng Zhang[1], Bo Li[3,*], and Guohong Ma[1,*]

[1]Department of Physics, Shanghai University, Shanghai 200444, China

[2]School of Physics and Electronics, Shandong Normal University, Jinan 250014, China

[3]Key Laboratory of Polar Materials and Devices (MOE), Department of Electronics, East China Normal University, Shanghai 200241, China

* To who correspondence should be addressed:

ghma@staff.shu.edu.cn (G. H. Ma), and bli@ee.ecnu.edu.cn (B. Li)

**Abstract:** Strong Coulomb interactions in two-dimensional systems, together with quantum confinement, make many-body processes particularly effective for carrier dynamics, which plays a crucial role in determining carrier lifetime, photoconductivity, and emission yield of the materials. Hereby, by using optical pump and terahertz probe spectroscopy, we have investigated the photocarrier dynamics in the $PdSe_2$ films with different thickness. The experimental results reveal that the photocarrier relaxation consists of two components: a fast component of 2.5 ps that shows the layer-thickness independence, and a slow component has typical lifetime of 7.3 ps decreasing with the layer thickness. Surprisingly, the relaxation times for both fast and slow components are exhibited both pump fluence and temperature independence, which suggests that synergistic interactions between defect trapping and Auger effect dominate the photocarrier dynamics in $PdSe_2$ films. A model involving defect-assisted Auger process is proposed, which can reproduce the experimental results well. The fitting results reveal that the layer dependent lifetime is determined by the defect density rather than carrier occupancy rate after photoexcitation. Our results underscore the interplay between Auger process and defects in two-dimensional semiconductors.

**Introduction**

Two-dimensional (2D) transition metal dichalcogenides (TMDs), the chemical formula $MX_2$ with M for transition metals and X for chalcogen, have attracted tremendous interest due to their unique optical and electrical performance and great application prospects in electronics and optoelectronics.[1-5] The ultrafast carrier relaxation dynamics in TMDs, including exciton dynamics, many-body interactions, defect trapping, *etc.*, have been studied in depth over the past a few years.[6-7] For example, the exciton-exciton annihilation (EEA), a four-body interaction, has been demonstrated in $MoS_2$ and $WS_2$, and the rate of EEA was calculated, which determines the upper limit of the quantum efficiency of 2D TMD-based light-emitting devices.[8-10] Li. *et al.* observed an anomalous photoluminescence quenching of multilayer $WSe_2$, and proposed that the carrier recombination in $WSe_2$ is dominated by Auger process, the similar effect had also been reported in $WS_2$,[11-12] which is of great significance to the photoconductivity response and emission yield of materials. Recently, due to the influence of inevitable defects during the materials growth on photocarrier dynamics, the interplay between defect-assisted and many-body interactions, such as exciton captured and Auger process, was studied in-depth.[13-15] All these provide a deep insight into the carrier dynamics in TMD and provide theoretical basis for their applications in optoelectronic devices.

As a member of 2D TMDs, palladium diselenide ($PdSe_2$), which exhibits the thickness-dependent electron bandgap,[16-17] higher carrier mobility,[18] and good air stability, has attracted great attention for its great potential in optoelectronic devices. In recent years, the optoelectronic performance and ultrafast dynamics of $PdSe_2$ have been further studied. As reported by Jia *et al.*, the large negative Kerr nonlinearity of $1.33 \times 10^{-15}$ $m^2$/W for $PdSe_2$ films, which is 2 orders of magnitude larger than that of bulk Si, indicating that $PdSe_2$ films have a unique advantage in high-performance nonlinear photonic devices.[19] Moreover, by using ultrafast optical pump-probe spectroscopy, the layer dependent interlayer breathing mode, as well as the longitudinal coherent acoustic phonon mode in $PdSe_2$ films have been studied, and the interlayer force constant and sound velocity are obtained in our previous study.[20] In general, the carrier dynamics in PdSe$_2$ and other TMD films are conventionally described as defect capture.[21-22] The detailed mechanisms, such as the roles of phonon, defect, and Auger process on the photocarriers, remain debate and to be elucidated.

In this study, we have investigated the photocarrier dynamics in layer dependent PdSe$_2$ films by using ultrafast optical pump-terahertz (THz) probe (OPTP) spectroscopy. Experimental results demonstrate that there are two time-scale for the dynamics of PdSe$_2$ films after photoexcitation: a fast component with time constant of ~2.5 ps and a slow one with time constant of ~7.3 ps. Surprisingly, both of the components show no significant dependence on the pump fluence as well as the temperature, which suggest that this was not simply a defect capture or multi-particle process, but more of synergistic interaction between the two processes. In fact, our results are consistent with mid-gap defect-assisted Auger effect. Two different time scales correspond to two different kinds of defects, *e.g.* fast and slow defect, respectively. Furthermore, a model involving defect-assisted Auger effect is proposed. The photocarrier dynamics and the occupancy of the carrier on the defect are obtained by solving the rate equation with the model, it is found that the layer-dependent lifetime change is caused by the defect density for different thickness films, and the lifetime change is irrelevant to the carrier occupancy rate with photoexcitation. Our study sheds light on the role of synergistic interaction between defect states and Auger effect on the ultrafast photoexcited carrier dynamics in PdSe$_2$ films, which opens up a new avenue for further optical performance development of PdSe$_2$ based devices.

**Methods**

The samples used in our experiment were PdSe$_2$ films with 5, 10, 15, and 30 layers, grown by chemical vapor deposition (CVD), and were transferred to sapphire substrate (provided by sixcarbon Tech, Shenzhen, China). The fabrication details and sample characterization are provided in reference 20. Figure 1a shows the crystal structure of PdSe$_2$, apparently, PdSe$_2$ has a puckered pentagonal atomic structure, in which each layer PdSe$_2$ consists of two planes of hexagonally arranged Se atoms linked to a hexagonal plane of Pd atoms via covalent bonds, and individual PdSe$_2$ layers are held together by weak van der Waals (vdW) forces. It is noted that the interlayer binding energies in PdSe$_2$ is as high as ~190 meV/atom,[16] resulting in the interlayer force constant is up to ~9.57 N/m,[20] so that the film shows highly robust stability in air, all of these make PdSe$_2$ a great potential application in optoelectronic devices.

Figure 1b illustrates the schematic diagram of the experimental setup. In our home-built ultrafast OPTP setup, the optical pulses are delivered from a Ti: sapphire amplifier with 120 femtoseconds (fs) duration at central wavelength of 780 nm (1.59 eV) and a repetition rate of 1 kHz. The laser pulse output from the amplifier is divided into three beams: the first two beams are used to generate and detect the THz signal, which constitutes the terahertz time-domain spectroscopy (THz-TDS) and is utilized to measure the static conductivity of the sample, in which the THz emitter and detector are based on a pair of (110)-oriented ZnTe crystals. The third beam acting as pump beam, is used to generate photocarrier in the samples. The pump induced THz transmission change (ΔT=T-T$_0$) with respect to delay time (Δt) was recorded, in which T and T$_0$ denote the transmission of the THz electric field peak value with and without pump beam, respectively. The optical pump and THz probe pulse are collinearly polarized with a spot size of 6.5 and 2.0 mm on the surface of sample, respectively. All measurements were conducted in dry nitrogen atmosphere, and the samples were placed in a cryostat with temperature varying from 5 K to 300 K.

The static THz conductivity for the 15L PdSe$_2$ displayed in Figure 1c was measured by implementing the THz-TDS without pump, and complex THz conductivity for 5, 10, and 30L PdSe$_2$ films are shown in Figure S1 of supplementary information (SI). For the very thin film of PdSe$_2$, Tinkham equation is applicable, and the complex conductivity can be fitted with Drude-Smith model:[23]

$$\sigma(\omega) = \frac{\omega_p^2 \varepsilon_0 \tau_s}{1 - i\omega\tau_s}\left(1 + \frac{c}{1 - i\omega\tau_s}\right), \quad \omega_p^2 = \left(\frac{n}{m_e}\right)\frac{e^2}{\varepsilon_0} \tag{1}$$

where ω$_p$ is the plasma frequency, ε$_0$ and τ$_s$ are vacuum dielectric constant and momentum scattering time, m$_e$ and n denote the electron effective masses (m$_e$ ~ 0.3 m$_0$ [24] with m$_0$ of free electron mass) and carrier density, respectively. The intrinsic carrier density of these films is estimated to be in the range of 0.85~2.4×10$^{20}$ cm$^{-3}$ according to eq 1. The high carrier density leads to strong Coulomb interactions in PdSe$_2$ films, which provide the possibility of many-body interactions such as Auger processes.[11-14, 25-26] At the same time, the inevitable defects in 2D TMD materials provide a platform for the interplay between Auger process and defects, the Auger processes for carrier capture by defects are believed to be possible, which consistent with our experimental results that will be discussed latter.

From the UV-visible-IR absorption spectra,[20] 1 μJ/cm$^2$ pump fluence at 780 nm is estimated to generate an electron (and hole) density of ~3.08~6.94×10$^{11}$ cm$^{-2}$ for 5~30L PdSe$_2$. And the measured THz transmission, $\Delta T/T_0$, can be expressed as,[13]

$$\frac{\Delta T}{T_0} \approx -2\eta_0 \frac{\Delta\sigma_r}{1+n_s} - 2\eta_0^2 \frac{(\sigma_r \Delta\sigma_r + \sigma_i \Delta\sigma_i)}{(1+n_s)^2} \quad (2)$$

where $\sigma_r$ ($\sigma_i$) is the real (imaginary) part of the optical conductivity of the sample, and $\Delta\sigma_r$ ($\Delta\sigma_i$) is the change of the real (imaginary) optical conductivity, e.g. the photoconductivity, $n_s \approx 3$ is the refractive index of the sapphire substrate, and $\eta_0$=376.7 Ω is the free-space impedance. Here, the effect of the second term on the right-hand side of eq 2 is negligible due to both and $\eta_0\sigma_i$ are much less than unity (*i.e.* $\eta_0\sigma_r(\eta_0\sigma_i) \ll 1$), $\Delta T/T_0$ is dominated by the first term on the right-hand side of eq 2, that is, magnitude of $\Delta T/T_0$ is dominated by the real part of photoconductivity $\Delta\sigma_r$, and $\Delta\sigma_r(\omega) \approx \left(\frac{\Delta n}{m_e} + \frac{\Delta p}{m_h}\right) \frac{e^2\tau}{1+\omega^2\tau^2}$.

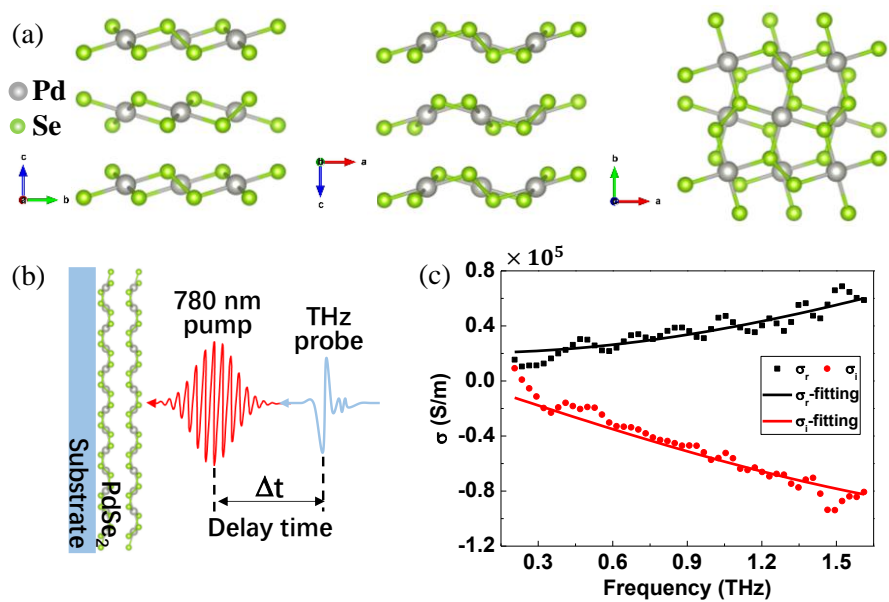
66666666666666666666666666666666666666666

Figure 1. (a) Crystal structure of PdSe$_2$, front view (left), side view (middle), and top view (right). The gray and green balls represent the Pd and Se atoms, respectively. (b) Schematic diagram of the optical pump-THz probe setup for studying photocarrier dynamics in PdSe$_2$ films. (c) The measured real and imaginary parts of the THz static conductivity of 15L PdSe$_2$ film along with the fitting curve by Drude-Smith model.

**Results and Discussion**

Figure 2a presents the normalized transient transmission, $\Delta T/T_0$ of 15L PdSe$_2$ film, with two selective pump fluence of 72 and 192 μJ/cm$^2$ at room temperature, more experimental data for 5, 10 and 30L PdSe$_2$ films are given in Figure S2 in SI. Obviously, the THz transient traces are identical under low and high pump fluence, which can be well fitted with biexponential function. Figure 2b shows the peak transmission with respect to pump fluence, and a linear pump fluence dependence is clearly seen. The fitting lifetime with respect to pump fluence is plotted in Figure 2c. Clearly, both the fast and slow lifetimes show no pump fluence dependence, in which the fast lifetime of $\tau_1$~2.5 ps and slow one of $\tau_2$~7.3 ps are obtained from the fitting. Figure 2d plots the amplitude ratio of $|A_1/A_2|$ with respect to pump fluence along with magnitude percentage of $A_1$ and $A_2$ (inset), it is clear that all the parameters including $A_1$, $A_2$ as well as $A_1/A_2$ show pump fluence independence.

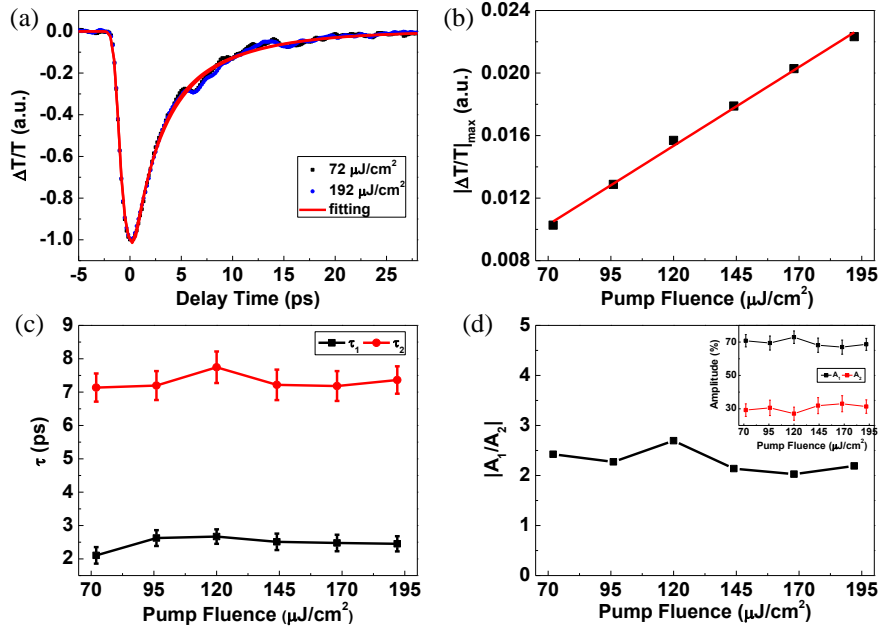

Figure 2. (a) Normalized transient transmission trace, $\Delta T/T_0$ of 15L PdSe$_2$ film under pump fluence of 72 μJ/cm$^2$ (black) and 192 μJ/cm$^2$ (blue) at room temperature along with biexponential fitting (red). (b) The maximum modulation depth with respect pump fluences. (c) The fitting lifetimes, $\tau_1$ and $\tau_2$ with respect to pump fluence. (d) The weight ratio ($|A_1/A_2|$) of two components with respect to pump fluence, the inset plots the pump fluence dependent the fitting amplitude $A_1$, $A_2$, respectively.

In order to further understand the photocarrier dynamics of PdSe$_2$ films, we also conducted OPTP measurement under different temperatures. Figure 3a shows the transient trace of $\Delta T/T_0$ with two temperatures of 5 and 300 K under pump fluence of 144 μJ/cm$^2$, and the inset plots the modulation of maximum $\Delta T/T_0$ (i.e. $|\Delta T/T_0|_{max}$) as a function of the delay time, apparently, the transient dynamics of PdSe$_2$ film does not change with temperature in the range from 5 K to 300 K. More experimental data are presented in Figure S3 in SI. Biexponential fitting can well reproduce the transient dynamics for PdSe$_2$ films, and the fitting lifetime with respect to temperature is presented in Figure 3b, in which the fast lifetime of ~2.0 ps and slow one of ~6.8 ps are obtained. Figure 3c plots the fitting amplitude ratio of $|A_1/A_2|$ with respect to temperature, and the inset shows the temperature dependence of the fitting magnitudes of $A_1$ and $A_2$, respectively.

Basically, the ultrafast photocarrier dynamic process consists of three main processes according to Figs. 2 and 3: i) After the photoexcitation at 780 nm, $\Delta T/T_0$ shows a negative change and reaches its maximum within 1 ps. ii) After that, a fast recovery of the $\Delta T/T_0$ occurs within 2~3 ps with a weight of approximately 70%. iii) Subsequently, a slow recovery lasts for 7~8 ps returning to the initial state. The photocarrier dynamics after photoexcitation follows biexponential relaxation function with fast one of $\tau_1$~2.5 ps and a slow one of $\tau_2$~7.3 ps. To our surprise, there are no significant pump fluence or temperature dependence for either fast or slow lifetime. Moreover, the weight ratio ($|A_1/A_2|$) of the two components, which also shows no obvious pump fluence or temperature dependence. Such interesting phenomenon imply that there are compound mechanisms that dominate the relaxation process, rather than a single one.

The possible mechanisms that can be responsible for prolonging or shortening the characteristic time of pump-induced carrier dynamics including phonon-assisted relaxation,[27-31] exciton–exciton annihilation (EEA),[8-9] Auger process,[12,32] and defect-assisted relaxation.[33-35] We first rule out phonon-assisted process, which depend strongly on the lattice temperature, since the energy released of the relaxation process needs to satisfy momentum conservation.[36] Our temperature independence of dynamics shown in Figure 3 suggests the phonon-assisted process is not important for both fast and slow relaxation processes. The probability of exciton-involved processes is expected to be much lower because the samples do not exhibit exciton absorption near the pump wavelengths used in our experiments. In contrast, the Auger rate of carrier capture is largely temperature independent, and TMD films fabricated with CVD method are known to contain many defects,[37] and defect-assisted Auger effect is proposed to be responsible to our experimental findings. PdSe$_2$ films exhibited no observable pump fluence and temperature dependence imply that there is synergistic interaction between defect-assisted and Auger process.

Our PdSe$_2$ films are n-type semiconductor that contain "fast" and "slow" defect states, and these defects are populated with electrons under equilibrium condition. After the illumination of the pumping pulse above the bandgap of PdSe$_2$ films, the photoexcited carriers in PdSe$_2$ films undergo a series of nonequilibrium processes, as illustrated in Figure 3d. Figure S4 in SI illustrates four basic Auger processes for electron and hole capture in PdSe$_2$ films. At first, electrons are jumped from valence band to conduction band by absorbing the energy of pump photons, numerous electron and hole are generated, and the electrons are populated at the bottom of conduction band and holes are populated at the top of valence band. Subsequently, the electrons at the "fast" defect states are recombined with the photogenerated holes at the top of valence band, and at the same time, the holes left in the "fast" defect states can trap the electrons at the bottom of conduction band, this process is a defect-assisted two-particle process, which corresponds to the fast component with typical lifetime of ~2.5 ps in our OPTP measurement. Additionally, the existence of "slow" defect states plays the similar way as that of the "fast" defect states, which is responsible for the slow component with typical lifetime of ~7.3 ps. Consider the density of the fast defect is higher than that of the slow one, therefore the fast process plays major role for the photocarrier dynamics. The TMD materials fabricated by CVD method are known to have many defects during the film growth process. It has been verified that there are many kinds of defects in CVD-PdSe$_2$ films, such as Se vacancies, surface defects caused by adsorption of oxygen molecules from the air.[37] The adsorption of O$_2$ at the film surface can be excluded based on the experimental data that the transient THz transmission are identical in air and vacuum (see Figure S5 in SI). It has been reported that there are two different defect states due to chalcogenide vacancies in TMD materials,[38] one possible source of fast (slow) defects in PdSe$_2$ films is two different defect states caused by Se vacancies.

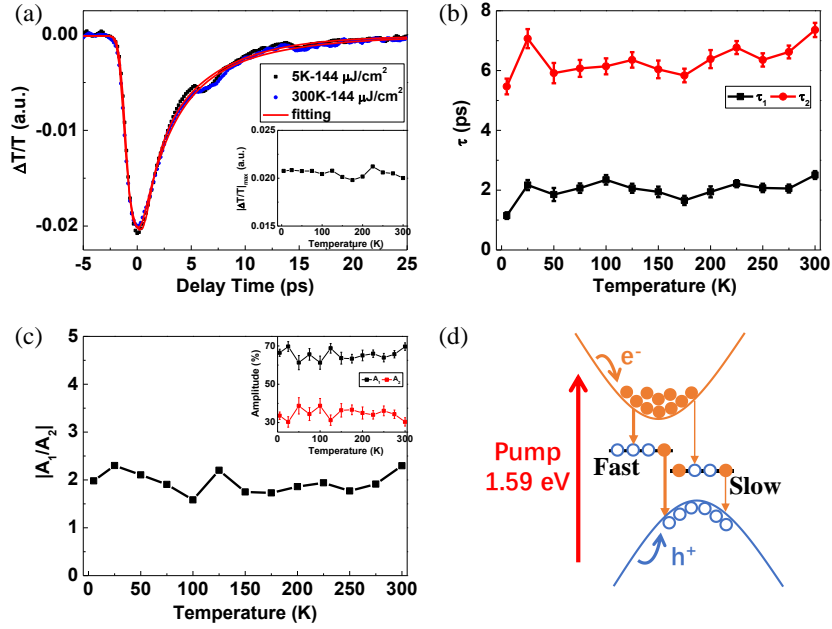

Figure 3. (a) The transient transmission $\Delta T/T_0$ of 15L PdSe$_2$ film under pump fluence of 144 μJ/cm$^2$ at 5 K (black) and 300 K (blue) along with biexponential fitting (red). The inset plots the maximum modulation depth with respect to temperatures. (b) The fitting lifetimes $\tau_1$ and $\tau_2$ with respect to temperature. (c) The weight ratio ($|A_1/A_2|$) of two components as a function of temperature, the inset is the fitting amplitude of two components $A_1$ and $A_2$ with respect to temperature. (d) Schematic for the synergistic interaction between defect trapping and Auger effect with above bandgap photoexcitation in PdSe$_2$ films.

In order to expound the impact of defects on ultrafast carrier dynamics in PdSe$_2$ films, we have carried out the OPTP measurement on PdSe$_2$ films with number of layers of 5, 10, 15, and 30L, respectively, which is shown in Figure 4a. Similarly, all experimental data can be well fitted with biexponential function as illustrated with the red line in Figure 4a, and the inset shows the maximum modulation depth with respect to the number layer of PdSe$_2$. The layer dependence of the fitting lifetimes, $\tau_1$ and $\tau_2$ are presented in Figure 4b. It is clear that the magnitude of $\tau_1$ is almost fixed at 2.2 ps, whereas that of $\tau_2$ decrease from 8.7 ps to 6.7 ps as the number of layers changes from 5L to 30L.

In order to consolidate our conjecture about the synergistic interaction between defect trapping and Auger effect, a model involving rate equations of synergistic interaction between defect trapping and Auger effect is used to fit all experimental data. The PdSe$_2$ films are n-type due to the existence of Se vacancies during the film growth process.[37] Here, assuming the defects are completely occupied in the thermal equilibrium state, and ignoring the emission of defects, the following rate equation can be obtained by only considering the Auger effect process in n-doped samples:[13]

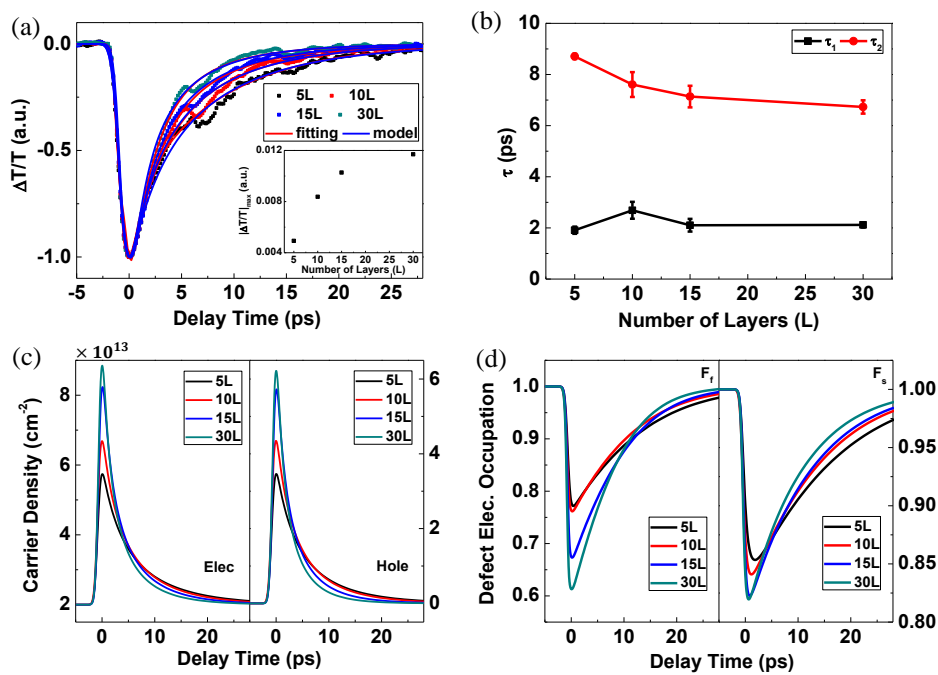

Figure 4. (a) Normalized transient transmission ΔT/T$_0$ (pump fluence normalized) of PdSe$_2$ films with number layer of 5, 10, 15 and 30 L under pump fluences of 144 μJ/cm$^2$, in which the dots are the experimental data, the red lines are the fitting curves with biexponential function, and the blue lines are the simulated results using the rate equations. The inset shows the maximum modulation depth with respect to the number of layers. The Auger carrier capture model reproduces all the time scales observed in the measurements over the entire range of the pump fluence values used. (b) Statistics of the relaxation time τ$_1$ and τ$_2$ for 5, 10, 15 and 30L PdSe$_2$ films. (c) The simulated electron and hole densities are plotted as a function of delay time after photoexcitation with pump fluence of 144 μJ/cm$^2$. (d) The simulated electron occupations of fast and slow traps, F$_f$ and F$_s$, are plotted as a function of delay time after photoexcitation with pump fluence of 144 μJ/cm$^2$

$$\frac{dn}{dt} = -A_f n_f n^2(1 - F_f) - A_s n_s n^2(1 - F_s) + gI(t) \tag{3}$$

$$\frac{dp}{dt} = -B_f n_f np F_f - B_s n_s np F_s + gI(t) \tag{4}$$

$$n_f \frac{dF_f}{dt} = A_f n_f p^2(1 - F_f) - B_f n_f np F_f \tag{5}$$

$$n_s \frac{dF_s}{dt} = A_s n_s p^2(1 - F_s) - B_s n_s np F_s \tag{6}$$

Here, *n* and *p* are the electron and hole densities, $n_f$ ($n_s$) is the density of fast (slow) defect levels, and $F_f$ ($F_s$) is the electron occupation of the fast (slow) defect level. $A_{f/s}$ ($B_{f/s}$) is the rate constant for the Auger effect of electrons (holes) by the defects. I(t) is the pump fluence in unit of µJ/cm². The consistency between $\Delta T/T_0$ simulation and experimental data for all PdSe$_2$ films are shown in Figure 4a, and the obtained parameters are presented in Table 1. It can be seen that the simulated curve with the model shows a good agreement with the experimental data as well as the biexponential function fitting. Through the above analysis, it can be concluded that the fast relaxation process $\tau_1$ is mostly contributed by the recombination of carriers in the fast defect, and a small fraction is contributed by the recombination of carriers in the slow defect. Therefore, $\tau_1$ is dominated by the first term on the right of eq 4, i.e. $n_f F_f$. The slow relaxation time $\tau_2$ is contributed by the slow defect capturing the remaining photoexcited electrons after all the holes are exhausted by the defects. Therefore, $\tau_2$ is dominated by the second term on the right of eq 3, i.e. $n_s(1-F_s)$. It is obvious that the fast (slow) relaxation lifetime $\tau_1(\tau_2)$ could be affected by the density of fast (slow) defect states $n_f(n_s)$ or by the electron occupancy on the fast (slow) defect states $F_f(F_s)$ or by both. In Figure 4c, the simulated electron (n, left) and hole densities (p, right) are plotted as a function of delay time after photoexcitation with pump fluence of 144 µJ/cm². And the simulated fast and slow defect electron occupancy $F_f$ (left) and $F_s$ (right) as a function of delay time are given in Figure 4d. It can be concluded that with the increase of the number of layers, the trend of $F_f(F_s)$ does not change significantly. On the other hand, as can be seen from the fitting parameter shown in Table 1, with the increase of the number of layers, the fast defect density $n_f$ is almost unchanged, while the slow defect density $n_s$ increases significantly, which is consistent with the result of biexponential lifetime fitting. Therefore, we concluded that the layer-dependent lifetime change is mainly dominated by defect state density.

Table 1. The Fitting Parameters obtained from Simulations

|  | 5L | 10L | 15L | 30L |
|---|---|---|---|---|
| $A_f$ (cm$^4$/s) | $1\times10^{-15}$ | $1\times10^{-15}$ | $1\times10^{-15}$ | $1\times10^{-15}$ |
| $n_f$ (1/cm$^2$) | $1\times10^{13}$ | $1\times10^{13}$ | $1\times10^{13}$ | $1\times10^{13}$ |
| $A_s$ (cm$^4$/s) | $3\times10^{-16}$ | $3\times10^{-16}$ | $3\times10^{-16}$ | $3\times10^{-16}$ |
| $n_s$ (1/cm$^2$) | $0.55\times10^{13}$ | $0.8\times10^{13}$ | $1.2\times10^{13}$ | $1.5\times10^{13}$ |
| $B_f$ (cm$^4$/s) | $5\times10^{-16}$ | $5\times10^{-16}$ | $7\times10^{-16}$ | $9\times10^{-16}$ |
| $B_s$ (cm$^4$/s) | $1\times10^{-16}$ | $1\times10^{-16}$ | $1\times10^{-16}$ | $1\times10^{-16}$ |
| g (1/μJ) | $4.43\times10^{13}$ | $5.8\times10^{13}$ | $8.5\times10^{13}$ | $10\times10^{13}$ |

**Conclusion**

In this study, we have investigated photocarrier dynamics in layer dependent PdSe$_2$ films by using optical-pump THz-probe spectroscopy. The photocarrier dynamics after photoexcitation show two different time scales: a fast lifetime of 2.5 ps and slow one of 7.3 ps, both of time constants show no apparent dependence on pump fluence and temperature, which suggests that the synergistic interaction between defect trapping and Auger effect dominates the photocarrier relaxation in PdSe$_2$ films. The Strong Coulomb interactions and inevitable defects in 2D materials make defect-assisted Auger processes more effective in PdSe$_2$ films. A model concerning defect trapping and Auger effect is used to fit the experimental results, the relaxation dynamics of electrons (holes) and the trapping of electrons (holes) by defect-assisted Auger processes are revealed. Our study of the photocarrier dynamics in PdSe$_2$ films pave the way for understanding and evaluating the performance of PdSe$_2$-based electronic and optoelectronic devices.

**Acknowledgements**: This work was supported by the National Natural Science Foundation of China (NSFC, Nos. 92150101, 61735010, and 61975110).